# Effects of Variation in System Responsiveness on User Performance in Virtual Environments

BENJAMIN WATSON[1], Univ. Alberta, Edmonton, Canada, and NEFF WALKER, WILLIAM RIBARSKY, and VICTORIA SPAULDING, Georgia Inst. Technology, Atlanta, Georgia, USA

System responsiveness (SR) is elapsed time until a system responds to user control. Over time SR fluctuates, so it must be described statistically with mean (MSR) and standard deviation (SDSR). This paper examines SR in virtual environments (VEs), outlining its components and methods of experimental measurement and manipulation. Three studies of MSR and SDSR effects on performance of grasp and placement tasks are then presented. The studies used within-subjects designs with 11, 12, and 10 participants, respectively. Results showed that SDSR affected performance only if it was above 82 ms. Placement required more frequent visual feedback, and was more sensitive to SR. We infer that VE designers need not tightly control SDSR, and may wish to vary SR control based on required visual feedback frequency.

[1]Requests for reprints should be sent to Benjamin Watson, Department of Computing, 615 GSB, University of Alberta, Edmonton, Alberta, Canada T6G 2H1.

Running Title: EFFECTS OF SYSTEM RESPONSIVENESS IN VR

# INTRODUCTION

Designers of virtual environments (VEs) face a fundamental tradeoff. Both a high level of detail (LOD, the visual complexity used in display) and good system responsiveness (SR, the time that elapses until the system responds to user control) are needed for good user performance. Unfortunately, achieving both high LOD and good SR is often not possible, since both goals require the same limited computational resources. This LOD/SR tradeoff has been identified as a critical issue facing the VE community (NSF, 1992; Van Dam, 1993).

One way to deal with this tradeoff is to actively vary the LOD (Funkhouser & Séquin, 1993) based on the need for different levels of SR. This would involve, for example, reducing LOD whenever the time required to render the model would result in a level of SR that impairs performance. However, unless the detail is precisely predicted and controlled, the system will fluctuate around the target level of SR. We will refer to this process as the management of LOD.

In order to control the LOD appropriately, one must first determine the capabilities of the system being used, in particular, optimal SR. Unfortunately SR is a very complex phenomenon, and is often misunderstood. This paper begins with an overview of SR, methods of measuring it, and ways in which it can be experimentally controlled.

LOD management also requires knowledge of how SR affects user performance in the system. In the context of an overview of SR, we present a review of existing research on this topic, and follow it with experiments examining the effects of SR on performance of two different tasks.

SYSTEM RESPONSIVENESS: MEASUREMENT AND MANIPULATION

VE systems typically use input devices to capture some element of real world state, such as hand position. Samples of this state are then represented in display. Even the fastest systems introduce a delay between the existence of a real world state and its eventual display. This delay has significant effects on system users, making it crucial that these effects be understood. SR and its related measures, outlined below, describe this delay.

We present our interpretation of SR in Figure 1. The figure shows three subsystems: a rendering system for display, a tracker system for sampling input, and the user. Time in the diagram moves horizontally from left to right.

Frame time is the amount of time a display sample (e.g. image) is shown on the display. (Frame time is closely related to frame rate, the number of samples displayed per second). In most VE systems, frame time is a multiple of video refresh time, since modern graphics displays synchronize image updates to the video screen refresh. These modern displays are also not interlaced.

System latency is the age of each sample presented on the display. For example, if a user's hand position is shown on the display, system latency describes how long it has been since the user's hand was actually in that position. System latency includes a portion of frame time (the exact amount varies from system to system) as well as the additional time required to collect an input sample from the real world (in Figure 1, the position of the tracker).

System responsiveness (SR) is the time elapsed from a user action until that action is displayed. SR is made up of system latency as well as the additional time between the completion of a user action and the next input sample used in display. More detailed discussions of the elements of SR can be found in (Watson, 1997), (Wloka, 1995) and (Mine, 1993).

---

Figure 1 about here

---

The terms system latency and SR are often used interchangeably, referring at times to the age of a sample, and at other times to the delay until feedback is received. This is unfortunate and can lead to much difficulty in interpreting published experiments. In this paper, the terms are not interchangeable. We focus in particular on SR, since we believe this time to feedback measure is the crucial to user performance.

It should be noted that the system presented in Figure 1 and the system used in our experiments were relatively simple systems, with only one display, one tracker, and one CPU. As the number of these different input and output elements in the system proliferates, the number of different frame times, latencies and levels of SR proliferates.

Frame time is relatively easily measured in software. Measuring system latency and SR is more complicated. System latency can be measured with a video camera viewing both a tracked object moving in a predictable manner (e.g. a pendulum) and the display presenting the tracked location of that object. Since the motion of the tracked object is predictable, measuring the distance between the tracked object and its represented location in a single frame of the video camera output measures the time elapsed since the tracked object was in the displayed location (system latency). Both Liang, Shaw and Green (1991) and Ware and Balakrishnan (1994) used this method of measurement in their experimental work. SR can also be measured with a video camera, this time by counting the number of video fields that elapse between an input event and the display of this event. This is the method used by Mine (1993) and in this paper. Note that frame time, system latency and SR will all vary randomly over time, with the degree of variation in system latency being greater than the degree of variation in frame time, and the degree of variation in SR being greater than the degree of variation in system latency. The pattern of this variation is complex and varies from system to system. It is important that not only the means of these display speed measures be recorded, but also their standard deviations if one is to be able to understand the effects of SR on user performance. Later in this paper we describe experiments that illustrate these measurements.

There are three software-based approaches for implementing experimental control of SR in VE systems. Each of them corresponds to different causes of change in SR over time.

Frame-latency manipulation varies frame time, system latency and SR simultaneously by adding delay between the receipt of an input sample at the displaying process, and the begin of display calculations. This mimics the quite common effects of fluctuations in model complexity, and resulting changes in the amount of computation required for display and input sensitive simulation. For example, adding delay between the receipt of a tracker sample and the rendering of the VE can mimic the delay that would result by rendering a higher LOD version of the same VE.

The second control approach, frame-only manipulation, varies frame time and SR, but not system latency. This can be done in the displaying process through addition of delay before the receipt of the input sample. This mimics the effects of changes in the amount of simulation computation not sensitive to input, or the amount of time already spent collecting other input samples. For example, adding delay before the receipt of a tracker sample can mimic the delay that would be introduced by increasing the accuracy of calculation for a real-time animation in the VE.

The third approach, latency-only manipulation, varies system latency and SR, but not frame time. This can be done by adding delay in an input management process running concurrently with the displaying process, and mimics the effects of fluctuations in the latency of input devices or the software that manages them. For example, if the tracker and display are connected to different computers, adding delay before any tracking data is sent to the displaying computer can mimic the delay that would result from improving position filtering on the tracker.

## HUMAN PERFORMANCE EFFECTS OF SYSTEM RESPONSIVENESS

There is a long history in psychophysics of the study of frame rate. Most of the research has focused on critical flicker frequency, the frame rate at which individual frames in a displayed series are no longer perceptible, and the series of displays is perceived as one continuous display. Critical flicker frequency varies with a number of variables, but generally does not exceed 75 Hz (Watson, 1986).

Unfortunately, the demands of generating images for display in VEs do not usually allow them to be displayed at frame rates quite so high as these. In the VE community, there have been a number of researchers who have used guidelines generated in the field to ensure adequate user performance. Airey, Rohlf and Frederick (1990) report that 6 Hz was an absolute minimum for their architectural walkthrough application. Pausch (1991) recommended 7 Hz, Card, Robertson and MacKinlay (1991), McKenna and Zeltzer (1992) and Bryson (1993) all recommend 10 Hz.

The long control delays in telerobotics have led researchers in the field to study the effects of frame-latency manipulation. Unfortunately, most of the studied latencies were well over a second, which might limit their meaning in the VE domain. Nevertheless, these studies (Ferrell, 1966) showed that latency can significantly impact task performance. Ranadive (1979) studied the tradeoffs of SR and visual detail in a remote manipulation task. He found a correspondence between performance and the number of bits of information delivered per second. It should be noted, however, that the environment being displayed in these experiments was largely static.

Bryson (1993) evaluated the effects of frame rate and latency in a 2D environment. Two participants performed tracking and placement tasks with frame-only and latency-only manipulation of mouse responsiveness. Results showed a linear relationship between tracking accuracy and both sorts of manipulation. At frame rates below 4 Hz and lags above 250 ms, the linear Fitts' law relationship between difficulty and performance time did not hold. MacKenzie and Ware (1993) performed a study of latency-only 2D placement with a larger number of participants. Latency affected performance significantly, and a modified (with latency and difficulty having a multiplicative relationship) Fitts' law proved effective at predicting placement times.

More recent studies have focused on the effect of system latency and SR using 3D

display technology. Tharp, Liu, French, Lai and Stark (1992) asked users to perform a highly demanding 3D tracking task. Users controlled a cursor with two table-mounted joysticks and viewed the VE with a head-mounted display (HMD) tracked only in two rotational degrees of freedom. With frame-only manipulation, performance reached asymptote when frame times fell below 100 ms. With latency-only manipulation, even 50 ms latencies harmed performance. Ware and Balakrishnan (1994) asked users to perform 3D placement tasks. Users viewed the VE with a fishtank system (head-tracked stereoscopic desktop display), and controlled a cursor with a 3D tracker. Placement accuracy and time were the dependent measures. In one experiment, participants moved the cursor until it was between two displayed planes. Latency-only manipulation of head motion responsiveness was not significant, but head motion was minimal in their system. Latency-only manipulation of hand responsiveness was significant. In another experiment, participants were asked to place the cursor inside a 3D box in three conditions: latency-only manipulation of hand responsiveness, frame-only manipulation of hand and head responsiveness, and frame-latency manipulation of hand and head responsiveness. Results did not show a strong linear Fitts' law relationship, especially at high difficulties. Ware and Balakrishnan speculated that this might be due to the difficulty of generalizing their Fitts' law based model from 2D to 3D.

There are a number of interesting trends in these studies. Frame-only manipulation seems to have weaker effects on placement and tracking than latency-only manipulation. Fitts' law is a useful predictor of placement performance, though less so when difficulty or the degrees of freedom in the task are high, or SR poor. Finally, there are indications that tracking is more sensitive than placement to manipulations of SR. This might indicate a relationship between the amount of feedback required by a task and SR.

## EXPERIMENTAL MOTIVATION

Our three experiments were motivated by practical questions that rose during our own design of VE applications. First, in our experiments participants remained standing while using an HMD to interact with the VE. Second, because we were interested in information useful for LOD management, we used frame-latency manipulation, duplicating the effects of change in visual detail (e.g. model complexity). As this detail changes, frame rates and SR fluctuate, and therefore we varied both mean SR (MSR) and standard deviation of SR (MDSR) during our studies. (A detailed discussion of the exact methods of measurement and control of these variables can be found in the Stimuli section below). None of the reviewed studies examined the effect of SDSR. Finally, we investigated the relationship between required task feedback and SR by using two task types. Knowledge of this relationship should prove very useful for dynamic control of LOD.

## EXPERIMENT 1

In the first experiment, we controlled MSR with levels of mean frame rate (9, 13 and 17 Hz) that are at the low end of rates typically found in current VEs. We also controlled SDSR with three frame rate standard deviations (0.5, 2.0 and 4.0 Hz). Participants performed tasks varying in the amount of visual feedback required (grasping and placement). We hypothesized that MSR and SDSR would have effects on the performance of both tasks.

### Method

Participants. Eleven undergraduate students participated in two 45 minute sessions. They were inexperienced in VEs and their vision was normal or corrected-to-normal (via contact lenses). The participants received course credit in an introductory course in psychology and were treated in accordance with APA guidelines. The participant with the best cumulative ranking at the end of the experiment received fifty dollars.

Apparatus. The experimental environment was displayed using a Virtual Research VR4 head-mounted display with Polhemus Isotrack 3D tracking hardware. The images were generated with a Silicon Graphics Crimson Reality Engine. The participants interacted with the VE using a plastic mouse, shaped like a pistol grip. During the experiment, they stood within a 1 m by 1 m railed platform. The platform was 15 cm high and the railing was 1.2 m high.

Stimuli. We controlled MSR and SDSR with frame-latency manipulation, which adds delay after the tracker is sampled to reach a targeted frame time. We measured SR in our system using the method outlined above and obtained a mean of 213 ms with a standard deviation of 30 ms in optimal conditions. According to both Wloka (1994) and Ware and Balakrishnan (1994), these levels are typical of current VEs.

Frame-latency manipulation of MSR may be implemented with levels described equivalently in frame time (e.g. 50 ms) or frame rate (e.g. 20 Hz). The same is not true of SDSR. Describing SDSR levels in frame time terms (e.g. ±10 ms) allows symmetric fluctuation around the MSR. However, as mean frame times decrease, the standard deviations that allow symmetric fluctuation also decrease. Describing SDSR levels in frame rate terms (e.g. ±4 Hz) is inherently asymmetric, with a bias toward longer SR times. Furthermore, the range of this fluctuation is dependent on the current frame rate mean. Since we were interested in the effects of large amounts of SR fluctuation even when MSR was excellent, and believed that the effects of this fluctuation would be primarily due to drops (not improvements) in SR, we chose asymmetric SDSR control with levels described in frame rate terms.

To control the pattern of the fluctuation in SR, we recorded a typical 218 frame sample of the frame times from an existing, uncontrolled and unmanaged VE application with approximately 7000 textured polygons. Frame time in the experiment was set by looping over this sample. Mean frame time was changed by finding the difference between the mean frame time in the original sample and the desired mean, and adding this difference to each sample. Frame time standard deviation was changed by scaling the adjusted sample around its new mean, effectively changing the range of its fluctuation.

The participants grasped a moving target object and placed it on a pedestal within a certain spatial accuracy tolerance (a box 29.5 cm in depth and width, 24 cm in height, for details see below). The target object was a white oblong virtual box, measuring 31 cm in height and 15.5 cm in depth and width. A yellow cubic cursor, 9 cm across each side, represented the mouse/hand location within the VE. The target object turned yellow and the cursor turned white when the participant successfully grasped the target object.

The VE consisted of a black floor with a white grid superimposed on it, and a black background. The target object traveled at a constant velocity of .75 m/sec from left to right along a circular arc of 125 degrees and 1.5 m in length, at a constant radius of 69 cm from the center of the platform. The ends of the arc were marked by tall white posts (see Figure 2). The target reached the end of the arc in 1.5 seconds and, after a 1.5 second pause, the target object reappeared at the left of the arc, effecting a wraparound. The target object moved up and down in an unchanging sinusoidal pattern. The amplitude of the sine wave measured 85 cm, with the bottom of the sine wave 1.3 m above the ground. The target object described exactly one period of the sinusoid before wraparound. The phase of the sinusoid was chosen randomly each time the target object appeared at the left end of the arc.

Figure 2 about here

The pedestal was white and located next to the base of the post marking the right end of the arc. It was an oblong box 1.5 m tall and 45 cm in depth and width. Success of the placement task was measured by testing the location of the target object: it had to be completely contained in a placement box. The placement box had the same depth and width as the pedestal and measured 55 cm in height. Since the target object was 31 cm tall and 15.5 cm in depth and width, it follows that placement was successful if it was oriented correctly and located within a 24 x 29.5 x 29.5 cm box. The placement box was blue and transparent and only appeared as feedback after the target object was incorrectly placed on the pedestal.

A red and white bullseye was centrally positioned on a solid black background between trials. Subjects could not begin a trial until they had centered this bullseye in their view.

When considering the importance of SR for human performance, it is important to consider the type of task being performed. One type of task decomposition that seems appropriate for describing interactions in a VE is the distinction between open-loop and closed-loop tasks (Wickens, 1992). Open-loop tasks are accomplished without the use of feedback during the task (e.g. throwing and jumping). In contrast, closed-loop tasks do use feedback, closing the feedback loop. Tasks of this type include driving and accurate tracking of moving objects. Of course, not all tasks are purely open- or closed-loop, but it is usually possible to locate a task at one end of this spectrum of feedback.

A trial consisted of the participant orienting on the bullseye and squeezing the trigger button on the mouse to begin a trial. After a random delay (between 750 and 1750 ms) the target object appeared, and the bullseye disappeared. To grasp the target object, participants had to squeeze the trigger button while the yellow cursor intersected the target object. Testing showed that the speed of the target object allowed only one grasp attempt. All participants in these studies adopted an open-loop “predict motion and intercept” grasping strategy, rather than a closed-loop “track motion and click” grasping strategy. When the target object was successfully grasped, it would shift to a location underneath the cursor. This made placement difficulty independent of grasp location. To complete the trial, the participant transported the target object to the right side of the visual field and placed it on the white pedestal. For the placement to be correct, the target object box had to be placed completely inside the placement box when the trigger was released. Participants required many corrective submovements to complete the task, and thus executed the majority of the placement task with closed-loop feedback.

Design. This study utilized a 3 (MSR controlled through mean frame rate) X 3 (SDSR controlled through frame rate standard deviation), within-subjects design. The levels used for frame rate mean and standard deviation are shown in Table 1, along with the corresponding values for MSR and SDSR.

Table 1 about here

There were four dependent measures, two for grasping and two for placement. Grasp time was the mean amount of time from the onset of target motion until the successful grasp, less any wraparound time when the object was invisible. Number of grasps was the mean number of button clicks from the onset of target motion until the successful grasp. Placement time was the mean amount of time from the successful grasp until placement was complete (the mouse button was released). Placement accuracy was the percentage of placements in which the target was placed within the placement box. Time measures included only correct trials (accurate placement within 30 seconds).

Placement time depended to a small extent on the location at which the grasp was made. However, the speed of the target object forced users to grasp in roughly the same location (the right of the arc), and most possible variation introduced in this way was captured in the grasp time analysis. Participants most often accomplished the grasp before the first wraparound.

Procedures. Each person participated in two sessions. Each session consisted of one block of 20 practice trials, followed by nine blocks of experimental trials, each block using a different experimental treatment. Each block began with three practice trials and was followed by five correct trials (accurate placement within 30 seconds, with 90 correct trials per participant over the two sessions). Participants were required to complete all trials within each display condition before ending the session. Block presentation order was varied randomly between participants and each order was used once.

Results

We analyzed our results in all experiments with two-way repeated measures analyses of variance (ANOVA) on the four dependent measures. Bonferroni pair-wise comparisons were performed to follow-up significant main effects. When there were significant interactions, they were analyzed with simple main effects tests. An alpha level of .05 was used for all analyses. In Table 1, the mean performance for the different dependent measures are presented. Results of the main ANOVAs are in Table 2.

Table 2 about here

Grasp Time. The effects of both MSR and SDSR were significant. Grasp time was longer when the MSR was controlled with a mean frame rate of 9.0 Hz than with a mean frame rate of 13.0 or 17.0 Hz. Grasp time was also longer when SDSR was controlled with a frame rate standard deviation of 4.0 Hz than with the 0.5 or 2.0 Hz levels.

There was a statistically significant two-way interaction between MSR and SDSR. The effect of MSR was significant at the 2.0 and 4.0 Hz SDSR frame rate standard deviations, but had no significant effect at the 0.5 Hz level. The effect of SDSR was significant only at the 9.0 Hz MSR mean frame rate.

Placement Time. Only MSR had a significant effect on placement time. Placement times at all levels of MSR differed significantly from one another.

Number of Grasps. Both MSR and SDSR had significant effects. Significantly more grasp attempts were made at the 9.0 Hz MSR mean frame rate than at the two higher mean frame rates. More grasps were made at the 4.0 Hz SDSR frame rate standard deviation than at the 2.0 Hz standard deviation, though neither of these standard deviations gave results different from those at the 0.5 Hz standard deviation.

The interaction of MSR and SDSR was again significant. MSR had a significant effect at the 4.0 Hz SDSR frame rate standard deviation, but not at the lower standard deviations. The effect of SDSR was significant at the 9.0 Hz MSR mean frame rate, but not at the higher levels.

Accuracy. Only the effect of MSR was significant, with the 9.0 Hz mean frame rate giving significantly less accurate results that the 17.0 Hz mean frame rate.

Discussion

These results show that both MSR and SDSR can affect performance. For both tasks, performance continued improving across the entire examined range of MSR (259 ms - 337 ms, see Table 1). SDSR affected only the grasping task, and only when this standard deviation was quite large (115 ms, or 4.0 Hz frame rate standard deviation, or, see Table 1). It appears that at poor levels of MSR, the prediction component of the grasping task is sensitive to the distribution in time of displayed visual samples.

The interactions observed between MSR and SDSR showed that effects were most significant when MSR was poor and SDSR high. This may be attributable to the increased

SDSR in these conditions that resulted from frame rate based control (see Table 1 and the discussion of stimuli above).

EXPERIMENT 2

This study had two goals. First, we wanted to examine the differential effects of improved levels of MSR on our two tasks. For the grasping task, results in the first experiment suggested that improving MSR further should have little effect on performance, and might eliminate the effects of SDSR. For the placement task, we expected continuing sensitivity to MSR, corresponding to the more frequent visual feedback required for the task.

Second, we wished to follow up the interactions of MSR and SDSR in the first experiment. Ideally, system designers should not have to consider the interaction of control of MSR and SDSR. We hypothesized that the interactions resulted from our frame rate based method of controlling SDSR. A new method of SDSR control was used in this experiment, one based on percentage of mean frame rate (e.g. 10% of each mean frame rate, rather than 2 Hz at all means). Since describing SDSR in this manner would decrease change in the range of SR fluctuation with MSR, we expected that SDSR would not interact with MSR.

**Method**

Below we note only the differences in methods between experiments 1 and 2.

Participants. There were twelve students recruited as participants in this experiment. None of these students had participated in the earlier experiment.

Design. Levels for MSR and SDSR differed from the first study (see Table 3). As outlined above, SDSR control was posed in percentage of frame rate mean, rather than absolute frame rate.

---

Table 3 about here

---

Results

We again applied the methods of analysis used in the first experiment. Below we discuss only significant results. Cell means are in Table 3, analysis results in Table 2.

Placement Time. The effects of both MSR and SDSR were significant. Placement times were longer when the MSR mean frame rate was 17 Hz than when it was 33.0 Hz. Placement times were shorter when SDSR was at 5.60% of mean frame rate than when it was at 44.40%.

Discussion

The effects of our manipulations corresponded well to our expectations. For the grasping task, improving MSR from 263 ms to 215 ms (mean frame rate from 17 to 25 Hz) did not have a significant effect, indicating a performance threshold. In addition, even the large ranges of fluctuation in SR introduced here had no effect. Thus the grasping task only proved sensitive to our manipulations at the poor MSR levels of the first experiment. In contrast, the placement task showed continued sensitivity to MSR, and a new sensitivity to SDSR. The placement task was more sensitive to SR than the grasping task.

Our predictions on SDSR control also proved accurate. SDSR was significant for placement even at superior levels of MSR, and the effect of SDSR showed no interaction with MSR. Controlling SDSR in terms of percentage of mean frame rate should allow simpler LOD management in VE systems.

## EXPERIMENT 3

In the final experiment, we sought confirmation of previous results. We expected that grasping would show little sensitivity to experimental manipulation even at slightly improved levels of MSR, while placement would show continued sensitivity to MSR. We also sought to test our belief that SDSR control in absolute frame rate terms is ineffective, because of its strong relationship to frame rate mean. In switching back to the absolute frame rate control of SDSR used in the first experiment, we expected little effect of SDSR on placement, because of the decrease in SR fluctuation at high frame rates. The interaction between MSR and SDSR exhibited in the first experiment would mostly likely be masked by this same effect.

### Methods

Below we note only the differences in methods between experiments 1 and 3.

Participants. Ten undergraduate students participated in this experiment. None had participated in either of the earlier experiments.

Design. Levels for MSR and SDSR differed from the first study (see Table 4). Absolute frame rate control of SDSR was used as in the first experiment.

---

Table 4 about here

---

### Results

We applied the methods of analysis used in the first experiment. Below we discuss only significant results. Cell means are in Table 4, analysis results in Table 2.

Grasp Time. The analysis of grasp time yielded one effect, SDSR, that approached significance ($F(2,18) = 3.07$, $p < .08$). This effect reflected a large increase in grasp time when the frame rate was 17 Hz and SDSR was most severe.

Placement Time. MSR had a significant effect. Placement time was longer with a frame rate of 17 Hz than with the two higher rates. The interaction of MSR and SDSR approached significance ($F(4,36) = 2.31$, $p < .08$).

Discussion

We received the confirmation we sought. Grasping was not affected significantly, while placement was affected by MSR. Absolute frame rate control again showed itself to be a poor method of SDSR control. Using this form of control eliminated SDSR effects at the high mean frame rates used in this experiment. The interaction of MSR and SDSR approached significance for placement, but was masked by the drop in the range of SR fluctuation at higher means.

The effect of SDSR on grasp time approached significance due to an outlier in the 17 Hz MSR mean frame rate, 7.8 Hz SDSR frame rate standard deviation condition. Subjects took far longer to perform in this condition in this experiment than in experiment 2. We believe this disparity in results occurred because experiment 2 had several conditions of high SDSR, while high SDSR occurred only in this condition in experiment 2.

## GENERAL DISCUSSION

This research has emphasized the distinction between SR and system latency, which are often confused in the VE field. We have reviewed elements of SR, methods for measuring it, and ways in which it can be experimentally manipulated. This review shows that SR fluctuates significantly over time, even if frame times are held absolutely constant. Moreover, changes in mean frame time can have a large effect, since SR includes more than one frame time.

Previous work showing differential effects of SR on tracking and placement tasks suggested a relationship between required task feedback and SR (e.g. Bryson, 1993; French, Lai & Stark, 1992; Ware & Balakrishnan, 1994). Our own experiments confirm this. Our open-loop grasping task was much less sensitive to MSR than our closed-loop placement task, which required frequent visual feedback. While performance on the grasping task reached a plateau as MSR improved, performance on the placement task continued improving up the limits of our experimental system.

Our results also show that SDSR can have a very significant effect on human performance. However, even for the fairly difficult tasks used in this study, SDSR must be quite large to have any sort of effect. In our experiments, SDSR of 82 ms or less was never significant. For the grasping task, SDSR was only significant at very poor MSR (in the 350 ms range). SDSR had no effect on the placement task at poor MSR. However, at improved levels of MSR (215 - 259 ms), the effect of SDSR was significant.

These results have a number of implications for VE and real-time graphics system designers. First, it is well-known that MSR affects user performance. However, since these effects vary with task type, designers may want to implement LOD management that is sensitive to the current task. In accordance with control theory research (Wickens, 1986), our results suggest that MSR requirements increase with the frequency of visual feedback needed in a task.

Second, users are sensitive to SDSR. However, SDSR must be quite severe before it affects user performance. This suggests control of SDSR can be quite loose. Our research indicates that posing SDSR control in percentage of mean frame rate is more effective than control in absolute frame rate terms. In related work (Watson, Spaulding, Walker & Ribarsky, 1997), symmetric fluctuation of SR had less effect than the fluctuation in this study, which was biased to longer SR. This suggests that transient improvements in SR should be of little concern to system designers; transient worsening of SR is more harmful to user performance.

There are a number of ways in which this research might be followed up. Our VE system was fairly simple, with limited parallelism. The field needs an examination of the effects and interactions of the many different levels of SR in highly parallel systems. We examined the continuum of required task feedback frequency with only two tasks; confirmation of our results with different tasks would be useful. Finally, implicit in the motivation of these studies is a tradeoff between LOD and SR. A direct experimental examination of the possible interactions of this tradeoff is certainly in order.

ACKNOWLEDGEMENTS

This research was funded in part by an Intel Fellowship.

**TABLE 1**

System Responsiveness and Measure Means in Experiment 1

| Mean Frm Rate | SdDv Frm Rate | Mean Frm Time | SdDv Frm Time | Mean Sys Resp | SdDv Sys Resp | Grasp Time (sec) | Place Time (sec) | Num Grasp | Acc Place |
|---|---|---|---|---|---|---|---|---|---|
| 9 | 0.5 | 111 | 6 | 337 | 60 | 3.58 | 3.72 | 2.61 | 84% |
| 9 | 2.0 | 117 | 26 | 345 | 82 | 3.30 | 3.72 | 2.10 | 85% |
| 9 | 4.0 | 133 | 55 | 370 | 115 | 6.32 | 4.08 | 3.68 | 78% |
| 13 | 0.5 | 77 | 3 | 285 | 47 | 2.90 | 3.36 | 2.18 | 87% |
| 13 | 2.0 | 79 | 12 | 288 | 57 | 2.37 | 3.40 | 1.86 | 87% |
| 13 | 4.0 | 85 | 30 | 298 | 77 | 2.82 | 3.38 | 2.21 | 93% |
| 17 | 0.5 | 59 | 2 | 259 | 41 | 2.46 | 2.99 | 2.05 | 85% |
| 17 | 2.0 | 60 | 7 | 260 | 46 | 2.05 | 2.77 | 1.89 | 95% |
| 17 | 4.0 | 62 | 15 | 263 | 55 | 2.26 | 3.04 | 1.95 | 96% |

**Frame rates are in Hz, frame times and system responsiveness are in ms.**

**TABLE 2**

Significant ANOVAs for All Three Experiments

| Exp | Dependent Measure | Exp Factors | Significant ANOVA Results |
|---|---|---|---|
| 1 | grasp time | MSR | $F(2,20) = 27.11, p < .001$ |
| 1 | grasp time | SDSR | $F(2,20) = 8.04, p < .01$ |
| 1 | grasp time | MSR x SDSR | $F(4,40) = 6.27, p < .001$ |
| 1 | place time | MSR | $F(2,20) = 22.75, p < .001$ |
| 1 | num grasps | MSR | $F(2,20) = 13.15, p < .001$ |
| 1 | num grasps | SDSR | $F(2,20) = 4.67, p < .05$ |
| 1 | num grasps | MSR x SDSR | $F(4,40) = 3.71, p < .01$ |
| 1 | accuracy | MSR | $F(2,20) = 3.93, p < .05$ |
| 2 | place time | MSR | $F(2,22) = 7.49, p < .01$ |
| 2 | place time | SDSR | $F(2,22) = 8.43, p < .01$ |
| 3 | place time | MSR | $F(2,22) = 9.32, p < .01$ |

**MSR is mean system responsiveness, SDSR is standard deviation of system responsiveness.**

**TABLE 3**

System Responsiveness and Measure Means in Experiment 2

| Mean Frm Rate | SdDv Frm Rate | Mean Frm Time | SdDv Frm Time | Mean Sys Resp | SdDv Sys Resp | Grasp Time (sec) | Place Time (sec) | Num Grasp | Acc Place |
|---|---|---|---|---|---|---|---|---|---|
| 17 | 5.6% | 59 | 3 | 259 | 42 | 2.21 | 1.80 | 1.58 | 93% |
| 17 | 22.2% | 62 | 14 | 263 | 54 | 3.13 | 2.12 | 1.93 | 91% |
| 17 | 44.4% | 58 | 140 | 257 | 189 | 3.02 | 2.20 | 1.83 | 86% |
| 25 | 5.6% | 40 | 2 | 230 | 36 | 2.52 | 1.80 | 1.74 | 94% |
| 25 | 22.2% | 42 | 10 | 233 | 44 | 2.36 | 1.95 | 1.73 | 87% |
| 25 | 44.4% | 39 | 95 | 229 | 128 | 2.22 | 2.08 | 1.48 | 85% |
| 33 | 5.6% | 30 | 2 | 215 | 33 | 1.98 | 1.72 | 1.48 | 90% |
| 33 | 22.2% | 32 | 7 | 218 | 38 | 2.51 | 1.88 | 1.69 | 93% |
| 33 | 44.4% | 30 | 72 | 215 | 103 | 2.54 | 1.99 | 1.81 | 92% |

**Mean frame rates in Hz, standard deviation of frame rates in percentage of mean frame rate, frame times and system responsiveness in ms.**

**TABLE 4**

System Responsiveness and Measure Means in Experiment 3

| Mean Frm Rate | SdDv Frm Rate | Mean Frm Time | SdDv Frm Time | Mean Sys Resp | SdDv Sys Resp | Grasp Time (sec) | Place Time (sec) | Num Grasp | Acc Place |
|---|---|---|---|---|---|---|---|---|---|
| 17 | 0.50 | 59 | 2 | 259 | 41 | 2.84 | 2.77 | 1.81 | 91% |
| 17 | 3.78 | 62 | 14 | 263 | 54 | 2.74 | 2.87 | 1.65 | 84% |
| 17 | 7.56 | 58 | 137 | 257 | 186 | 4.21 | 2.94 | 2.15 | 84% |
| 33 | 0.50 | 30 | 1 | 215 | 32 | 2.61 | 2.81 | 1.67 | 90% |
| 33 | 3.78 | 30 | 4 | 215 | 35 | 2.88 | 2.43 | 1.92 | 94% |
| 33 | 7.56 | 31 | 5 | 217 | 36 | 2.76 | 2.22 | 1.69 | 89% |
| 41 | 0.5 | 24 | 1 | 213 | 30 | 2.24 | 2.27 | 1.57 | 92% |
| 41 | 3.78 | 25 | 2 | 213 | 31 | 2.61 | 2.39 | 1.66 | 85% |
| 41 | 7.56 | 25 | 5 | 213 | 34 | 2.72 | 2.42 | 1.87 | 84% |

**Frame rates are in Hz, frame times and system responsiveness are in ms.**

LIST OF FIGURES

**Figure 1.** The components of system responsiveness in a simple single processor VE system, consisting of three parallel components: user, tracker and renderer. Here both system latency and system responsiveness exceed frame time.

**Figure 2.** A top down schematic of the experimental environment. Users on the platform begin by looking at the bullseye; the target object moves left to right across the visual field.

BENJAMIN WATSON, NEFF WALKER, WILLIAM RIBARSKY and VICTORIA SPAULDING (Effects of Variation in System Responsiveness on User Performance in Virtual Environments)

BENJAMIN WATSON received his B.S. degree in Computer Science from UC Irvine and his M.S. and Ph.D. degrees in Computer Science from the Georgia Institute of Technology. He recently took a faculty position with the Department of Computing Science at the University of Alberta. His research interests are in novel human-computer interfaces, computer graphics, and virtual environments.

NEFF WALKER received his undergraduate degree from Princeton University and his Ph.D. in cognitive psychology from Columbia University. Since 1991 he has been at the Graphics, Visualization and Usability Center at the Georgia Institute of Technology. His research interests are advanced human-computer interfaces, aging, and movement control. He is currently working for UNAIDS in Geneva.

WILLIAM RIBARSKY received a Ph.D. in physics from the University of Cincinnati. He is Associate Director for Service and Computing at the Graphics, Visualization and Usability Center. His research interests are data visualization, collaborative computing, and virtual reality.

VICTORIA SPAULDING received her undergraduate degree from the University of Illinois at Urbana-Champagne. She is in the doctoral program at the Georgia Institute of Technology's School of Psychology, where she is funded by a fellowship from AT&T Laboratories. Her research interests are human-computer interaction, virtual environments, and aging.

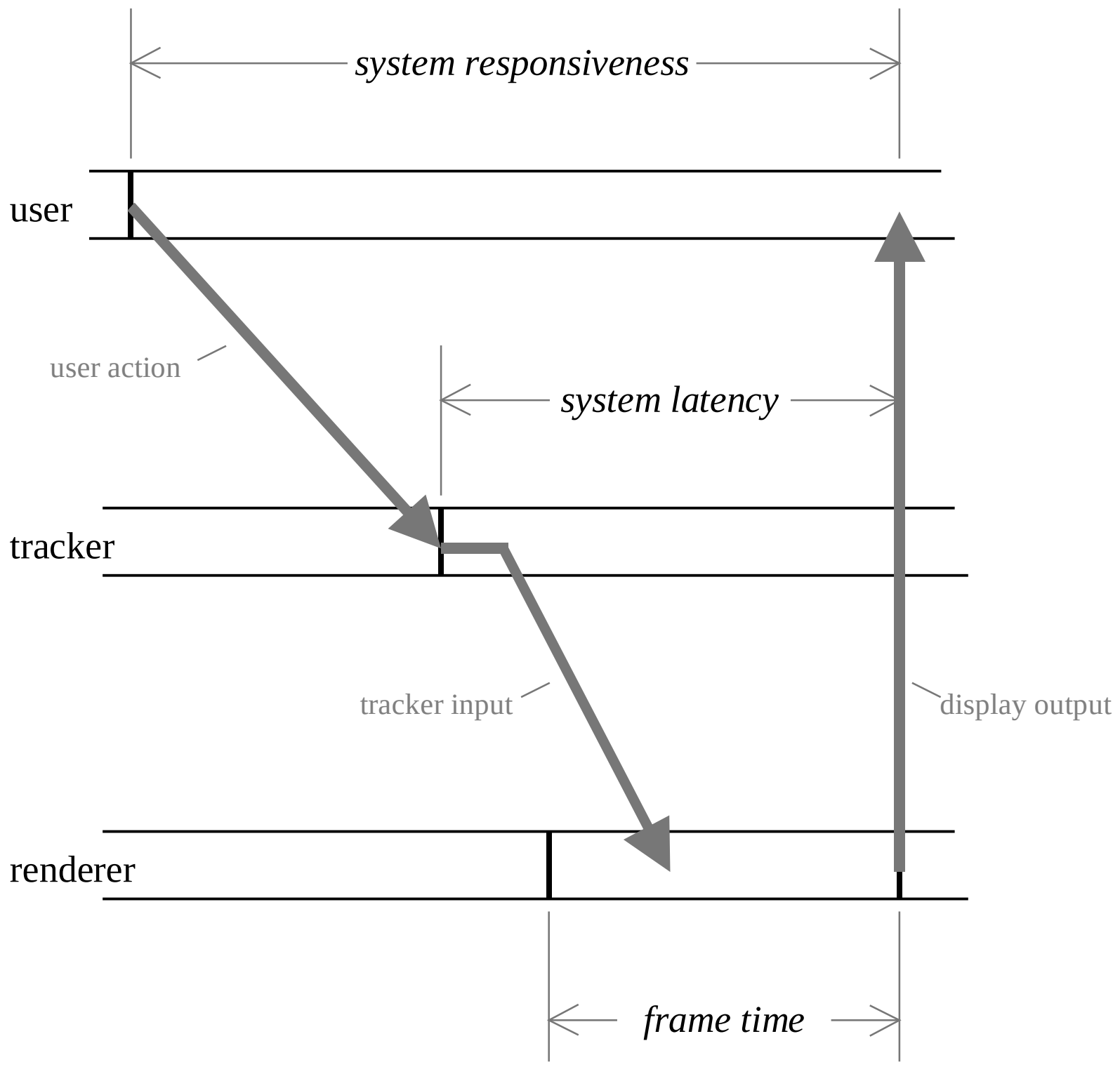

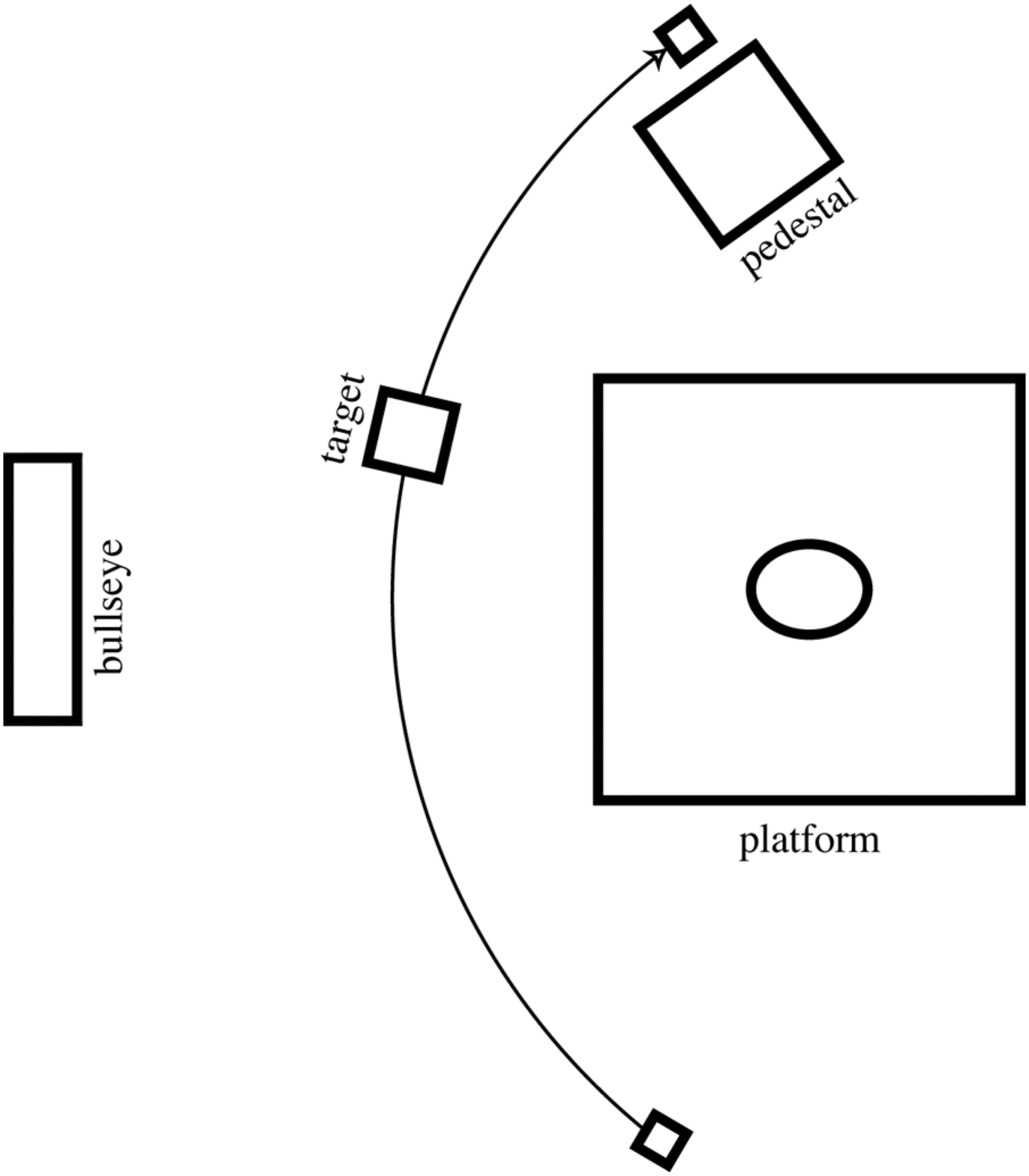
pedestal
target
bullseye
platform